\documentclass[aps,prl,superscriptaddress,showpacs,twocolumn]{revtex4}
\usepackage{graphicx}
\begin{document}
{
\title{Photon-assisted Fano Resonance and Corresponding
Shot-Noise in a Quantum Dot}
\author{Zhongshui Ma}
\affiliation{State Key Laboratory for Mesoscopic Physics and
Department of Physics, Peking University, Beijing 100871, China}
\affiliation{Advanced Research Cneter, Zhongshan University,
Guangzhou 510275, China }
\author{Yu Zhu} \affiliation{State Key
Laboratory for Mesoscopic Physics and Department of Physics,
Peking University, Beijing 100871, China}
\author{Xin-Qi Li}
\affiliation{NLSM, Institute of Semiconductors, CAS, Beijing
100083, China}
\author{Tsung-han Lin}
\affiliation{State Key Laboratory for Mesoscopic Physics and
Department of Physics, Peking University, Beijing 100871, China}
\author{Zhao-Bin Su}
\affiliation{Institute of Theoretical Physics, Chinese Academy of
Science, Beijing 100080,China}

\begin{abstract}
We have studied the Fano resonance in photon-assisted transport in
a quantum dot and calculated both the coherent current and
spectral density of shot noise. It is predicted, for the first
time, that the shape of Fano profile will also appear in satellite
peaks. It is found that the variations of Fano profiles with the
strengths of nonresonant transmissions are not synchronous in
absorption and emission sidebands. The effect of interference on
photon-assisted pumped current has been also investigated. We
further predict the current and spectral density of shot noise as
a function of the phase, which exhibits an intrinsic property of
resonant and nonresonant channels in the structures.
\end{abstract}


\maketitle

\medskip

The phenomena of photon-assisted tunneling (PAT) have been
intensively studied in a variety of mesoscopic systems. The main
feature of PAT is that the monochromatic radiation induces
additional tunneling processes when electrons exchange energy by
absorbing or emitting photons, and the conductance displays a
strong nonlinearity and develops
sidebands\cite{kouwenhoven2,kouwenhoven1}. However, the previous
studies of PAT have not considered the possibility of symmetric RT
being violated in structures where the nonresonant transmission is
taken part in the tunnelings between two reservoirs. Recently,
G\"ores {\it et al}.\cite{gores} and Zacharia {\it et
al}.\cite{zacharia} demonstrated in their experiments the RT
processes involving the interference with the nonresonant
tunneling (NRT) channel. The phenomenon has stimulated much
experimental\cite{madhavan} and theoretical\cite{stone,schoeller}
interest. The coexisting of two possible tunneling channels
(resonant and nonresonant), interfering with each other, through
the single-electron transistor shows prominent perspectives for
the Fano effect\cite{fano}, i.e., asymmetric line shapes (ALS) in
conductance. Because the interference between energetically
accessible RT channels and NRT channel has not been investigated.
In this sense, it is interesting to ask whether the interference
exhibits ALS on satellites peaks and what the characteristic
features are manifested in PAT.

On the other hand, the spectral density of current noise,
especially its nonequilibrium aspect (so-called shot noise),
reveals very useful information hidden in current
measurements\cite{buttiker1,iannaccone}. The study on
photon-assisted shot noise (PASN) is developed very fast in recent
years. Lesovik and Levitov\cite{lesovik} modelled an open single
connected loop threaded by a time-varying Aharonov-Bohm flux to
study the flux-controlled interference. This offers a clever way
of affecting its dynamics and enables one to study PASN. PAT on
nonequilibrium current fluctuation has been observed by Schoelkopf
et al. experimentally\cite{schoelkopf}. To our interest, here, a
new degree of freedom is provided by the employment of NRT channel
in tunneling processes. The correlation between RT and NRT
processes is high in time for transmitting electrons. The
investigation of spectral density of shot noise in PAT has twofold
significations. On the one hand, the study of this quantity can be
used as a cross-check for the analysis of the interference of the
charge transferring, in the case of an existence of continuum
level degenerate in energy with a discrete resonant level. On the
other hand, the investigation on the zero bias PASN, in which the
time average current vanishes, enables us to clarify intrinsic
effect of discrete charges in interference with Fano resonance.

In this Letter, we present the theoretical analysis of Fano
resonance in PAT and demonstrate those interesting features for
the ALS on satellite sidebands. Our analysis is based on a small
configuration so called a quantum dot (QD) connected to two
electronic reservoirs via tunneling barriers. To capture the
essential physics of photon-assisted Fano resonance we consider
the simplest case with noninteracting electrons in a single-level
QD. The NRT channel is incorporated by a bridge channel as a kind
of direct tunneling between the continuum states of two
reservoirs, which can be thought as a point-like-contact added to
the configuration. This Letter consists of two parts: First, we
study the Fano effect in PAT and reveal the characteristics of ALS
in peaks located on satellites; Second, we investigate the
spectral density of shot noise in photon-assisted Fano resonance.

The system is described by a tunneling Hamiltonian
$H=H_{leads}+H_d+H_{ldr}+H_{lr}$, in which
$H_{leads}=\sum_{k\alpha}\epsilon_{k\alpha}(t)
c^\dagger_{k\alpha}c_{k\alpha}$, $H_d= \epsilon_d(t)d^\dagger d$,
$H_{ldr}= \sum_{k\alpha}(t_{k\alpha}
c^\dagger_{k\alpha}d+t^*_{k\alpha} d^\dagger c_{k\alpha})$, and
$H_{lr}= \sum_{kk'}(v_{kk'}c^\dagger_{kl}c_{k'r}+v^*_{kk'}
c^\dagger_{kl}c_{k'r})$, where $c^\dagger_{k\alpha}$
($c_{k\alpha}$) creates (annihilates) an electron of momentum $k$
in reservoir $\alpha$ ($=l$, $r$); $d^\dagger$ ($d$) is the
creation (annihilation) operator for an electron in QD;
$t_{k\alpha}$ are tunneling coefficients that describe the
tunneling between the reservoirs $\alpha$ and the quantum dot,
while $v_{kk'}$ describes the NRT between the reservoirs. The
Hamiltonian is time dependent due to energy shifts,
$\epsilon_{k\alpha }(t)=\epsilon_{k\alpha}+eV_\alpha\cos\omega_1
t$, of the reservoirs and illuminated QD by irradiation,
$\epsilon_d(t)=\epsilon_d+eV_\omega\cos\omega_2 t$. Using
formulism of time-dependent Green's functions\cite{wingreen} the
current flowed out the left lead is $I_l(t)=(2e/\hbar)Re
[\sum_{kk'}v_{kk'}G^<_{kr,k'l}(t,t)+
\sum_{k}t_{kl}G^<_{d,kl}(t,t)]$, where $G^<_{d,kl}$ and
$G^<_{kr,k'l}$ are Keldysh Green functions. In the current the
term involving $G^<_{kr,k'l}$ contains the NRT and its
interference with RT while the term involving $G^<_{d,kl}$
contains the RT and its interference with NRT. These
nonequilibrium Green functions can be evaluated by employing the
Keldysh technique and be expressed in terms of Green function
$G_{dd}$ at QD and the bare Green functions in the leads. The
connection among them in a set of Dyson-like equations includes
all possibilities of scattering tunneling and interference. For
simplicity, we neglect the energy dependence of various couplings
and write the intrinsic linewidth in terms of the density of
states, $\rho_\alpha$, in the leads $\alpha$ under the wide
linewidth approximation, i.e.,
$\Gamma_\alpha=2\pi\rho_\alpha|t_\alpha|^2$ ($\alpha=l,r$) and
$W=(2\pi)^2\rho_l\rho_r|v|^2$. Besides, it should be emphasized to
point out that there is a mix coupling among the left and right
barriers, and the bridge channel over the two reservoirs, i.e.,
$\Delta=(2\pi)^2\rho_l\rho_rt_lv^*t^*_r=\sqrt{\Gamma_l\Gamma_rW}e^{i\varphi}$.
It is found that the combination of $t_lv^*t^*_r$ can be regarded
as an existence of intrinsic accesses constructed by RT and NRT
channels and gives rise to a phase dependence. The phase $\varphi$
manifests its contribution throughout the following calculations
and shows an Aharonov-Bohm-like effect. The time average current
is obtained as $I=(e/\pi\hbar)\int
d\epsilon\sum_{mn}f_{mn}(\epsilon) T_{mn}(\epsilon)$ with a
definition of $f_{mn}(\epsilon)=
J^2_n(u_\omega)[f_l(\epsilon)J^2_m(u_l)-f_r(\epsilon)J^2_m(u_r)]$,
where $J_n(u_\beta)$ is the $n$th-order Bessel function of
$u_\beta=eV_\beta/\hbar\omega_\beta$ ($\beta=\omega$, $l$ and $r$)
and $f_\alpha (\epsilon)$ is the Fermi distribution function of
electrons in the reservoir $\alpha$. $T_{mn}$ describes the
transmissions of the tunneling through both RT and NRT channels,
the interference between them, and all multiple scattering on the
dot and the contacts in the processes of absorption and emission
of photons. The analytic formulation is given by
\begin{equation}
T_{mn}(\epsilon)={T_b\epsilon^2_{mn}+{1\over
4}{\tilde\Gamma}^2\Omega+\epsilon_{mn}{\tilde \Gamma}^2
\sqrt{T_b\Omega}\cos\varphi\over(\epsilon_{mn}+{1\over
4}\Gamma\sqrt{T_b\Omega}\cos\varphi )^2+{1\over
4}{\tilde\Gamma}^2}
\end{equation}
where $\epsilon_{mn}=\epsilon-\epsilon_d+m\omega_1-n\omega_2$,
$\Gamma=\Gamma_l+\Gamma_r$, ${\tilde \Gamma}=\Gamma
T_b/2(1-\sqrt{1-T_b})$, and $\Omega^2=4\Gamma_l\Gamma_r/\Gamma^2$.
We leave $T_b=W/(1+W/4)^2$ as a turnable parameter, so-called
transparency of NRT channel, in our following discussions. This is
equivalent in controlling the NRT strength. It is a transmission
coefficient in the quantum point contact\cite{yeyiti}.
Ref.\cite{yeyiti} has discussed in details that $T_b$ varies
between zero and one as a function of $W$. In eq.(1) the
renormalized levels are given by
$E_{mn}=2[\epsilon_d-m\omega_1+n\omega_2-\Gamma\sqrt{T_b\Omega}\cos\varphi/4]/{\tilde
\Gamma}$ and show satellite resonant levels offset from the main
nonlinearity by the shifts corresponding to absorption and
emission of photons. If we express them in terms of resonant
scattering phase shifts $E_{mn}=cot\delta_{res}^{mn}$, it is
predicted that the sidebands (with the resonant scattering phase
shifts in different integer multiples of $\omega_1$ and
$\omega_2$) are still remained and but with a shift due to a self
energy renormalized the resonance
$\Gamma\sqrt{T_b\Omega}\cos\varphi/4$. According to the discussion
of Hofstetter et al.\cite{schoeller}, eq.(1) can be written as a
generalized Fano form with an asymmetric parameter,
$q=\sqrt{\Omega(1-T_b)/T_b}\cos\varphi$ which is independent of
PAT. The common Breit-Wigner line shapes of PAT is recovered at
$q\to\infty$ ($T_b=0$), whereas $q\to 0$ ($T_b=1$) yields a
symmetric antiresonance. ALS would be explicitly noted at the
finite value of $q$. Because of resonant accesses depending on
exchange energy of photons, it is predicted that a Fano zero
associated with each PAT resonance results in the dips (ALS) on
the satellite peaks of current characteristics.
\vspace{-0.25 cm}
\begin{figure}[ht]
\centerline{\includegraphics[width=8.6cm,angle=0]{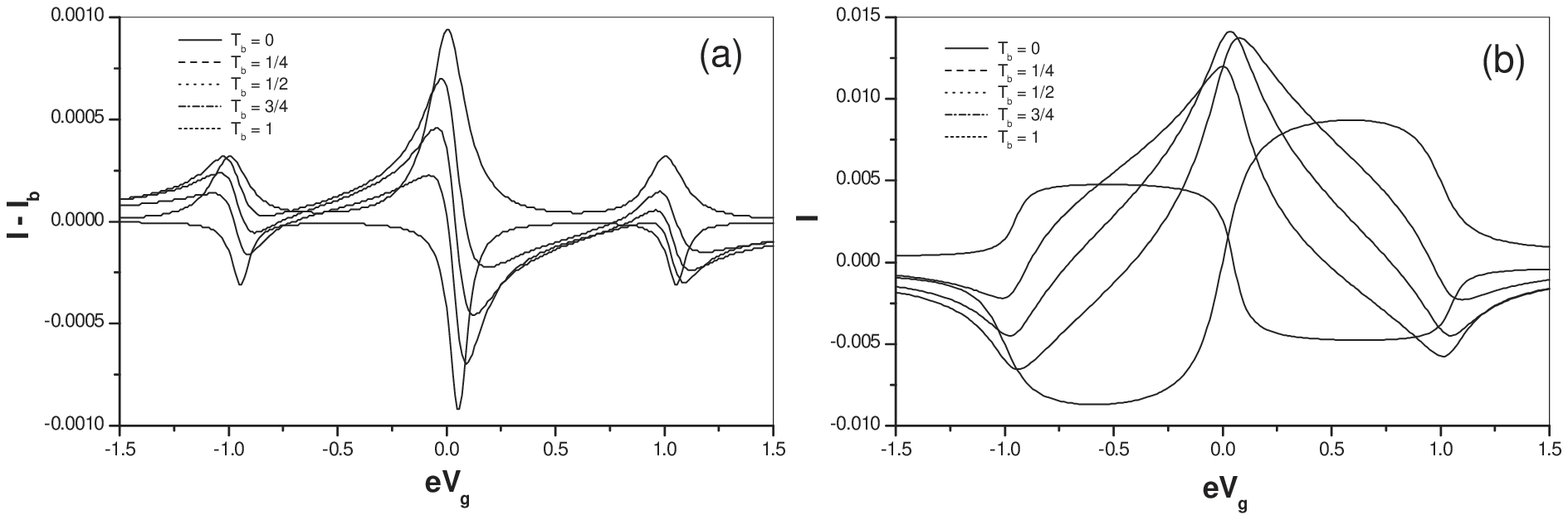}}
\vspace{-0.25cm} \caption{Current as a function of $eV_g$ for
$T_b=$0, 0.25,0.5,0.75, and 1. The parameters are $u_\omega=0$,
$\omega_1=1$, $\varphi=0$, and dc bias $V=0.01$. (a) the symmetric
irradiation on the both leads, $u_l=u_r=1$, and (b) the
photon-assisted pumped current under asymmetric illumination,
$u_l=1$ and $u_r=0$. $I_b$ is the current at
$\Gamma_l=\Gamma_r=0$. } \label{Fig.1}
\end{figure}
Fig.1 displays the current at zero temperature as the function of
$eV_g$, the gate voltage, where the resonant level in QD has been
taken as an energy reference, i.e., $\epsilon_d=0$. Throughout
this Letter, the energy is measured in unit of $\omega_1$
($e=\hbar=1$) and the symmetric tunneling barriers
($\Gamma_l=\Gamma_r=0.1\omega_1$) are considered. From Fig.1a we
find that the satellite peaks are produced at $eV_g=n\hbar\omega$
as expected. The amplitude of any given sideband is a
non-monotonic function of photon flux in the form of a nontrivial
dependence on the Bessel function $J^2_n(u_\alpha)$ and the
interference. Fano line shapes on the sidebands imply that the PAT
accesses interfere with the NRT channel. The influence of the Fano
resonance has been also represented in Fig.1a by running $T_b$
over the regime from the weak NRT to the one that NRT contribution
to conductance is comparable in size to the RT component.
Increasing $T_b$ the current $T_{mn}$ makes the line shape run
away from Breit-Wigner line shape ($T_b=0$) and exhibits a
Breit-Wigner dip in the limit when NRT dominates the tunneling
($T_b=1$). The dips on the line shapes located on satellites
saturate at the same finite values symmetrically at the limitation
attributed in the NRT channel. It is interesting to notice that
the unsynchronized asymmetrizations on the satellite peaks located
two sides of main resonance are visible in varying $T_b$. Such
unsynchronous dips in the absorption and emission of photons can
been explained as follows: The scattering cross-section of
tunneling is proportional to a function of sum $m$ over
$\sin^2(\delta_{res}^{m}+\delta)$\cite{schoeller}, where $\delta$
is a nonresonant phase shift related to the asymmetric parameter
$q$, i.e., $q=-cot\delta$. As pointed above the resonant
scattering phase shifts $\delta_{res}^{m}$ are different for the
processes of absorption and emission of photons. Thus the definite
value $T_b$ ( thus $\delta$) results in the different values in
$\delta_{res}^{m}+\delta$ for the accesses of emitting photons
(positive $m$) and absorbing photons (negative $m$).
Alternatively, eq.(1) is a sum of two even and one odd functions
of $\epsilon$ around its poles. The part of odd function
asymmetrizes the line shape and leads to Fano profile. The third
term of the numerator in eq. (1) indicates that the processes of
absorption and emission of photons affect the Fano profiles in a
different weight. It is implicit in the variations of ALS with the
strengths of nonresonant transmissions are not synchronous in
absorption and emission sidebands. This can be understood as that
the tunneling electrons will acquire different coherent phases
from the stimulated absorption and emission respectively, which
would be exhibited in the interference between RT and NRT
channels.

It is well known that an asymmetric illumination in the reservoirs
reveals photon-assisted pumping. Without NRT, Kouwenhoven {\it et
al}.\cite{kouwenhoven1} has studied the photon-assisted pumping at
zero dc bias voltage. We set $V_r=0$ and a finite $V_l$. The
photon absorption occurs only at the left reservoir, which leads
to the net current to the right when the RT level above the Fermi
level of reservoirs while reverses when RT level below the Fermi
level of the reservoirs. Without NRT the pumped current changes
sign when the gate voltage is swept such that the RT level moves
cross the Fermi level of reservoirs. However, it is not truth when
NRT is joined. These processes interfere with NRT shows a wide
region of current to the right. The pumped current occurs over a
width corresponding to the photon energy. It is noticed that the
pumping current is reversed at $T_b=1$ with respect to the case of
$T_b=0$. In Fig.1b we have represented several possibly pumped
current profiles with different values of $T_b$. Note that except
for the limiting situations $T_b=0$ and $T_b=1$, the profiles are
asymmetric. There are two zero-current points instead of only one
for the case of limiting situations. This behavior originates from
an incomplete suppression due to the interference near the
degeneracy point of single-particle energies. As one can see in
eq. (1), current indicates the Fano line shape is phase dependent
and a periodic function of $\varphi$ with period 2$\pi$. The
symmetry in $\varphi$ and $-\varphi$ refers to as ``phase
locking'', which is an exact property of two-terminal
setups\cite{buttiker1}. Fig.2 demonstrates the current-phase
relation for $T_b=0.25$ and $1.00$. From Fig.2a it is evident to
note the property that there is a symmetry $I(\varphi, eV_g)$ and
$I(\varphi-\pi, -eV_g)$, which follows from the symmetries in
electron and hole. It is also interesting to notice that the
sidebands are smeared away at $\varphi=\pi/2$ (see in Fig.2b)
because the Fano parameter vanishes at $\varphi=\pi/2$. It can
been seen that we have the usual FAT resonance for $T_b=0$ and the
disappearance of interference for $T_b=1$ at $\varphi=\pi/2$.
\begin{figure}[ht] \centerline{\includegraphics[width=8.0cm]{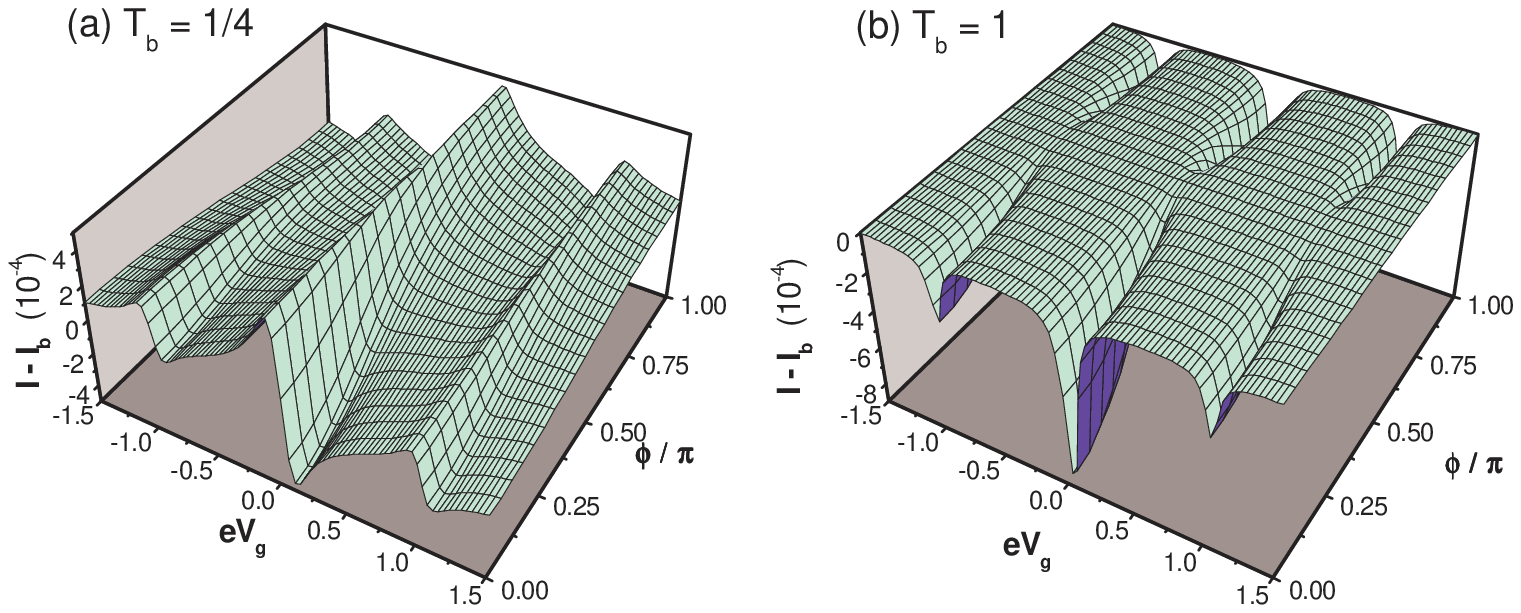}}
\vspace{-0.25 cm} \caption{Current as a function of $eV_g$ and
$\varphi$ for the symmetric illumination for $T_b=$ (a) 0.25 and
(b) 1. Other parameters are the same as those given in Fig.1a}
\label{Fig.2}
\end{figure}

To learn more about the influence offered by the interference with
NRT we study the spectral density of shot noise $S(\omega)$ and
the effect in affecting its dynamics. It is defined as $S(\omega)
=\int d(t-t')\exp [i\omega (t-t')]<\delta {\hat I}(t)\delta {\hat
I}(t')+\delta {\hat I}(t')\delta {\hat I}(t)>$, where $\delta
{\hat I}(t)={\hat I}(t)-<{\hat I}(t)>$ is the fluctuation in
current. In general the noise in PAT oscillates in time with the
frequencies of external microwave radiations. $<\delta {\hat
I}_\alpha(t)\delta {\hat I}_\beta(t')>$ ($\alpha,\beta=l,r$) does
not need to have the same values for different pairs of
$\alpha\beta$ in the presence of external irradiation. We
calculate $S(\omega)=\sum_{\alpha\beta}S_{\alpha\beta}(\omega)$
instead of $S_{\alpha\beta}(\omega)$. Because of $<{\hat
I}_l(t)>=-<{\hat I}_r(t)>$, $S(\omega)$ calculated here relates to
the zero-point current fluctuation. Using the technique of
nonequilibrium Green function as done in calculation of current it
makes analytical studies possible. However, the calculation is
more complicated because of the correlation among the electrons in
the difference processes taking into account. Taking the notions
${\tilde\omega}_{mn}=\omega-\epsilon-
\epsilon_d+m\omega_1-n\omega_2$ and ${\cal
F}_{\beta}(\omega,\epsilon,\epsilon')=
(1-f_\beta(\epsilon))\delta(\omega-\epsilon'+\epsilon)-
f_\beta(\epsilon)\delta(\omega-\epsilon'-\epsilon)$, we obtain the
spectral density of shot noise,
$S(\omega)=(e^2/\pi\hbar^2)\sum_{\alpha mn}\int d\epsilon
f_\alpha(\epsilon)J^2_m(u_\alpha)J^2_n(u_d)[s^{mn}_{\alpha}(\omega,\epsilon)+\sum_{\beta}
\int d\epsilon' J^2_m(u_\beta) J^2_n(u_d) {\cal F}_\beta
(\omega,\epsilon,\epsilon')s^{mn}_{\alpha\beta}(\omega,\epsilon,\epsilon')]$,
where
\begin{equation}
s^{mn}_{\alpha}(\omega,\epsilon)={\Gamma_\alpha
{\tilde\Gamma}\over ({\tilde\omega}_{mn}+{1\over
4}\Gamma\sqrt{T_b\Omega}\cos\varphi )^2+{1\over
4}{\tilde\Gamma}^2}
\end{equation}
\begin{eqnarray}
s^{mn}_{\alpha\beta}(\omega,
\epsilon,\epsilon')={[{\tilde\Gamma}_\alpha+{1\over 4}
W{\tilde\Gamma}_{\bar\alpha}+(-{1\over 2})^\alpha
\Gamma\sqrt{T_b\Omega}\sin\varphi]{\tilde\Gamma}_\beta\over
({\tilde\epsilon}_{mn}+{1\over 4}\Gamma\sqrt{T_b\Omega}\cos\varphi
)^2+{1\over 4}{\tilde\Gamma}^2}
\end{eqnarray}
\begin{figure}[ht]
\centerline{\includegraphics[width=6.0cm]{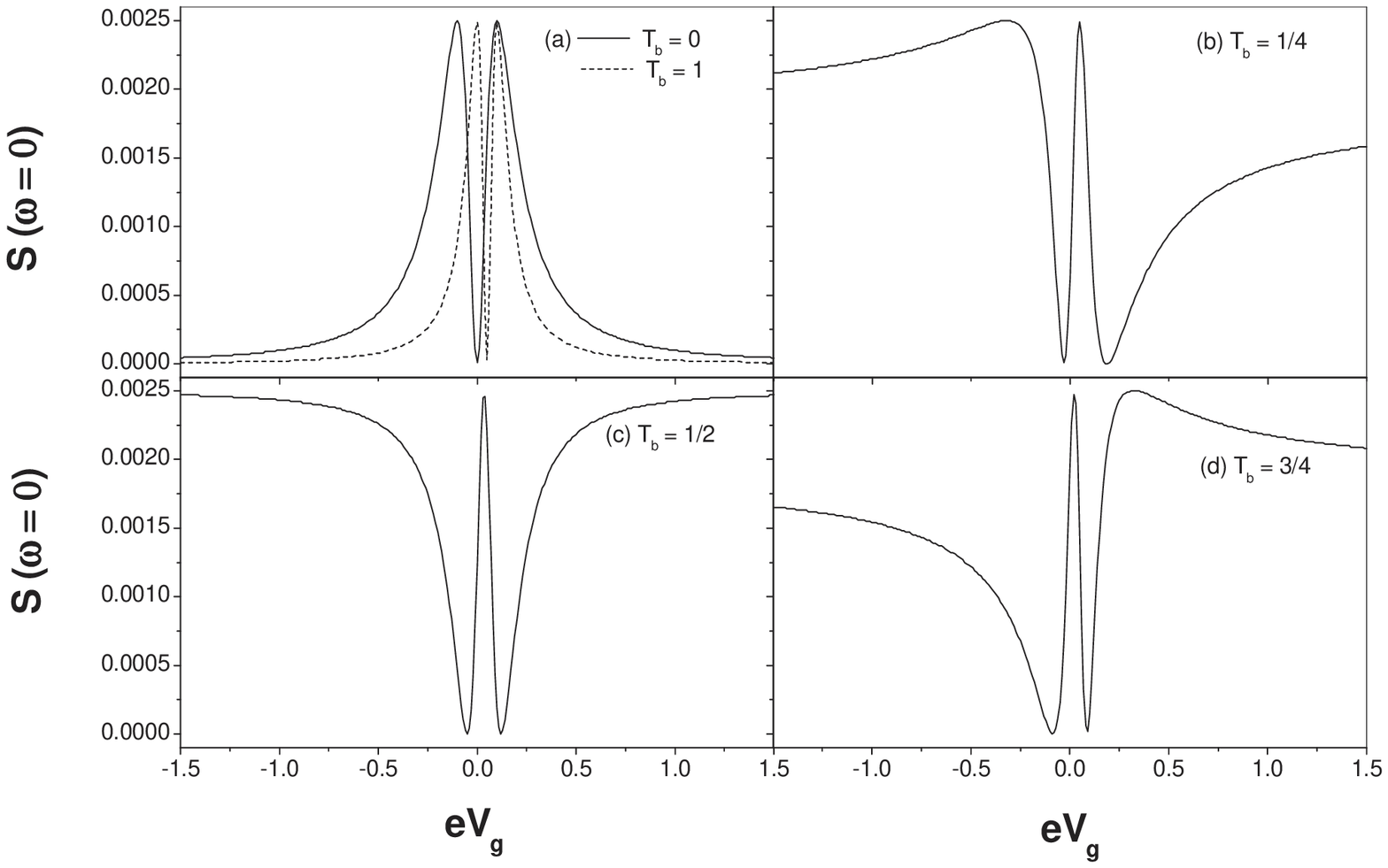}}
\vspace{-0.25cm} \caption{Spectral density of shot noise as a
function of $eV_g$ in absence of irradiation
($u_l=u_r=u_\omega=0$) for $T_b=$ 0, 0.25, 0.5, 0.75, and 1;
$\varphi=0$ and dc bias $V=0.01$.} \label{fig.3}
\end{figure}

The dc and zero-frequency spectral density of shot noise at a
finite dc bias and zero temperature as the function of gate
voltage is plotted in Fig.3. It contains a new information as
compared with the current in the evolution of the transparency of
NRT, $T_b$. By analyzing the asymmetric variation of shapes with
varying the transparency, $T_b$, it is found that the spectral
density of shot noise recovers its original form at $T_b=1$ as
that at $T_b=0$ and reverses shape at $T_b=0.5$. The result is
different from what has been seen in the current.
\begin{figure}[ht]
\centerline{\includegraphics[width=8.0cm]{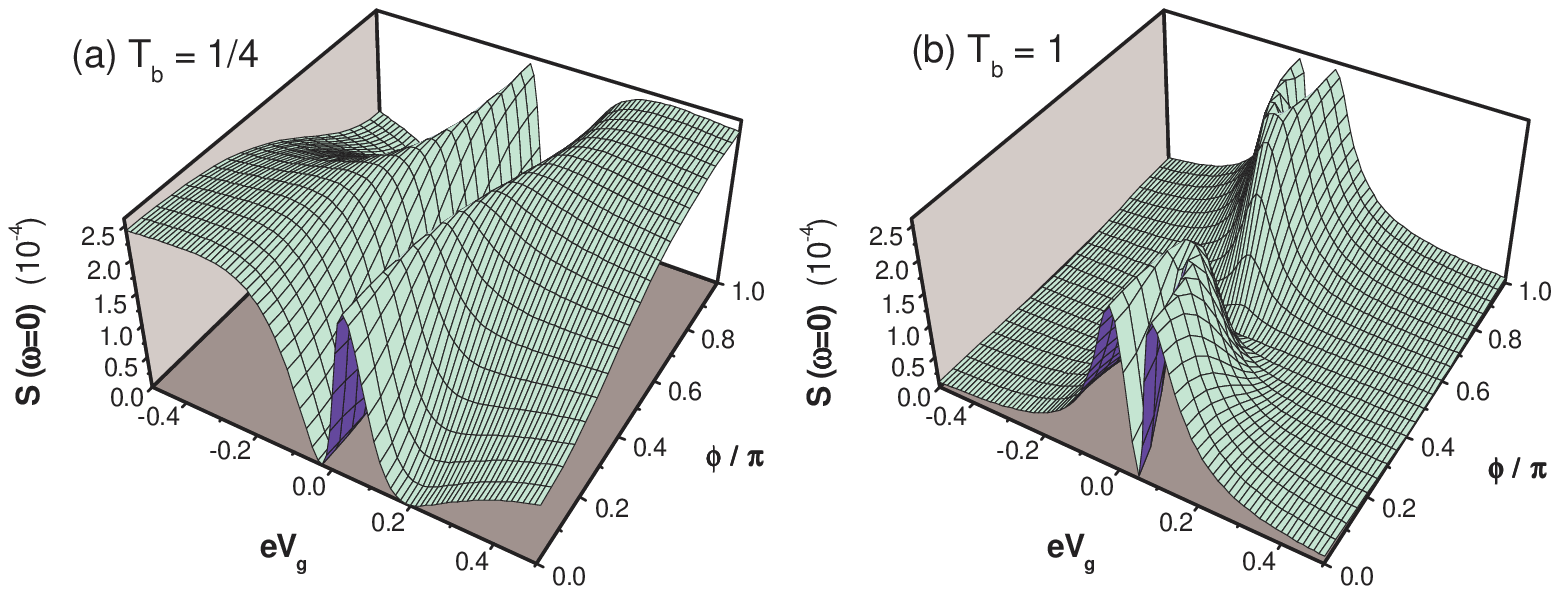}}
\vspace{-0.25cm} \caption{Spectral density of shot noise as a
function of $eV_g$ and $\varphi$ for the case in Fig.3 at $T_b=$
(a) 0.25 and (b) 1.} \label{fig.4}
\end{figure}
The phase-dependence has been illustrated in Fig.4. This is a
robust feature which survives for all Fano resonant tunneling in a
structure of QD. It is an unambiguous signature of the fact of
Aharonov-Bohm-like oscillation. The difference between here and
the description of Lesovik and Levitov\cite{lesovik} is that the
phase dependence is not an external flux but the intrinsic
existence. The shot noise remains the constant at $\varphi=\pi/2$
and $T_b=1$ in Fig.4b indicates the interference smearing over
fluctuation and leaving noise as a constant background. This is
agreement with that we have observed in the current in Fig.2b. The
fact reveals the symmetry of phase dependence is the intrinsic
property and is independent of the absorption and emission of
photons.

Before making adieu we would like to avail ourselves of a little
space to face an interesting question: what the photon-assisted
signature in the shot noise may be and what effect does
nonresonant channel give risen in? To this we plot the spectral
density of shot noise at zero bias voltage in Fig.5. Because no
time-average current exists when the ac voltage drop is the same
for both reservoirs\cite{kouwenhoven1} the fluctuations on the
current are completely induced by the external radiations. From
the plot of shot noise in Fig.5, the predicted modulation respect
to the absorbtion and emission of photons does arise. It is seen
that a continuing increase of transparency of NRT channel, the
varying on the top of peaks in the zero-bias spectral density of
shot noise does also reveal the asymmetric Fano resonance. It
indicates the possibilities of relaxation among these states
excited by absorbtion and emission of photons. An electron absorbs
a photon or emits a photon and hops to a energetically accessible
above the Fermi level in the leads. The interference with the
forward and backward possibility in the nonresonant channel leads
to the ALS in the zero-bias shot noise. The phase dependence of
zero-bias shot noise also shows the fact that the interference
smears over fluctuation and leaves noise as a constant background.
\begin{figure}[ht]
\centerline{\includegraphics[width=4.5cm]{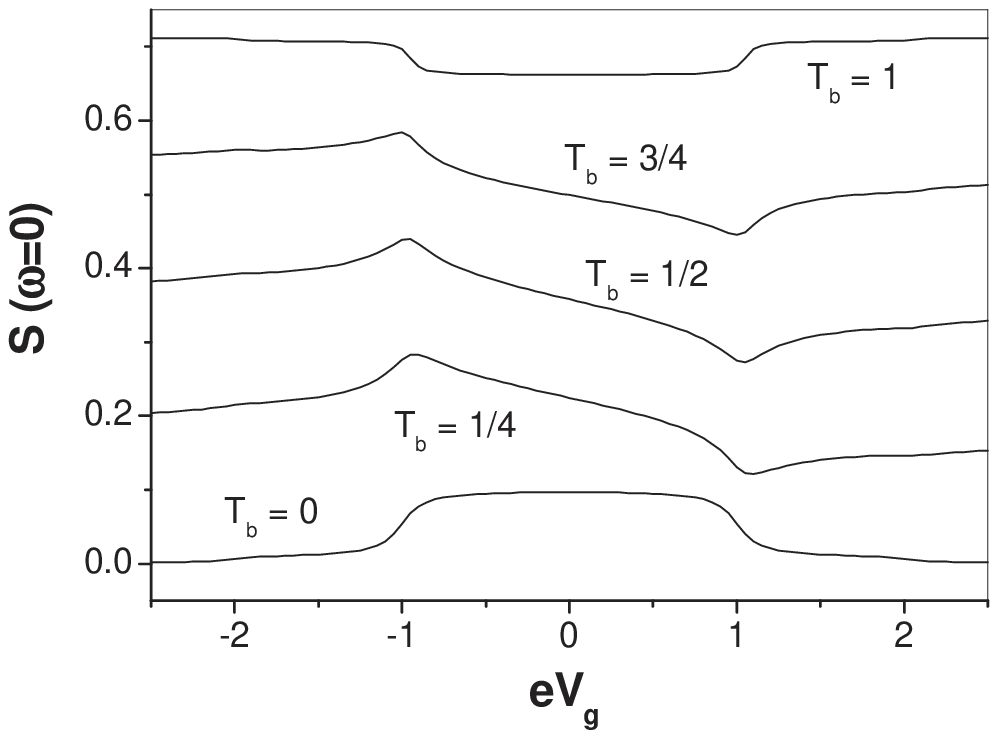}}
\vspace{-0.25cm} \caption{Spectral density of shot noise as a
function of $eV_g$ in present of irradiation for $T_b=$ 0, 0.25,
0.5, 0.75, and 1 at zero bias voltage. The parameters are
$\varphi=0$, $u_\omega=0$, and $u_l=u_r=1$.} \label{fig.5}
\end{figure}

In summary, we have demonstrated that the shape of Fano profile
appears in peaks located on absorption and emission sidebands. The
asymmetrization of peaks on those satellites, corresponding to
absorbing and emitting photons respectively, are not synchronous
in varying the strengths of nonresonant transmissions. Both the
current and spectral density of shot noise depend on a phase which
is essentially a sore of intrinsic phase difference between the
energetically accessible RT and NRT channels.
\begin{acknowledgments}
We acknowledge the support from NNSFC and NSF of Guangdong
Province under Grant No. 90103027, 10274069, and 011151.
\end{acknowledgments}

\end{document}